\documentclass[
aps, 
prd, 
twocolumn, 
superscriptaddress, 
floatfix, 
nofootinbib 
]{revtex4-1}

\usepackage{ifpdf}
\ifpdf
\usepackage{cmap} 
\usepackage[pdftex]{color, graphicx} 
\usepackage{epstopdf} 
\usepackage[pdftex, colorlinks=false, pdfstartview=FitH, hypertexnames=false, unicode, bookmarks=false]{hyperref} 
\else
\usepackage[dvips]{color, graphicx} 
\usepackage[hypertex, colorlinks=false, unicode]{hyperref} 
\fi

\graphicspath{{Pics/}{Pics/More/}}

\usepackage{amssymb, amsmath, amsxtra}

\begin{document}

\title{
Influence of magnetic field on beta-processes in supernova matter
}

\author{Alexandra Dobrynina}
\email{dobrynina@uniyar.ac.ru}
\affiliation{P.\,G.~Demidov Yaroslavl State University,
	Sovietskaya 14, 150003 Yaroslavl, Russia}

\author{Igor Ognev}
\email{ognev@uniyar.ac.ru}
\affiliation{P.\,G.~Demidov Yaroslavl State University,
	Sovietskaya 14, 150003 Yaroslavl, Russia}


\begin{abstract}
An influence of a magnetic field on beta-processes is investigated under conditions of a core-collapse supernova. For realistic magnetic fields reachable in astrophysical objects we obtain simple analytical expressions for reaction rates of beta-processes as well as the energy and momentum transferred from neutrinos and antineutrinos to the matter. Based on the results of one-dimensional simulations of a supernova explosion, we found that, in the magnetic field with the strength $B \sim 10^{15}$~G, the quantities considered are modified by a few percents only and, as a consequence, the magnetic-field effects can be safely neglected, considering neutrino interaction and propagation in a supernova matter.
The analytical results can be also applied for accretion discs formed at a merger of compact objects in close binary systems.
\end{abstract}

\maketitle

\section{Introduction}

Core-collapse supernova (SN) is the final stage of a star evolution with a mass $M_{\rm star} \gtrsim 10 \, M_\odot$. For the first time, an importance of neutrinos for the SN was recognized by Colgate and White~\cite{Colgate:1966ax} and Arnett~\cite{Arnett:1966}. A recent era of a study of neutrino effects in collapsing stars has begun after the foundation of the unified electroweak theory of fundamental interactions~--- the Weinberg-Salam-Glashow model~\cite{Tanabashi:2018oca}. 
Neutrinos play a significant and sometimes even dominant role in all phases of the SN explosion. The $\beta$-processes, also called direct URCA-processes:
\begin{align}
p + e^{-} &\to n + \nu_e\,, \label{eq:p1} \\ 
n + \nu_e &\to p + e^{-}\,, \label{eq:p2} \\ 
n + e^{+} &\to p + \bar{\nu}_e\,, \label{eq:p3} \\ 
p + \bar{\nu}_e &\to n + e^{+}\,, \label{eq:p4}
\end{align}
are the dominant neutrino processes in a SN matter~\cite{Bruenn:1985en}.
They provide an energy exchange between neutrinos and the matter and change a chemical composition of a matter. Note an absence of the $\beta$-decay, $n \to p + e^- + \bar\nu_e$, because it is kinematically suppressed in the SN matter. 

An origin of a magnetic field with a strength up to $10^{15}$~G, extracted from observables from magnetars~\cite{Olausen:2013bpa} and other classes of isolated neutron stars~\cite{Vigano:2013lea}, is one of the hot topics of modern astrophysics. Several more or less successful models have been proposed to explain its appearance. 
As shown in Refs.~\cite{Heger:2004qp,Peres:2018mww,Kissin:2018}, the magnetic field strength in the iron core at the pre-supernova stage is $B \sim (10^9 - 10^{10})$~G that yields the magnetic field strength of order of $B \sim (10^{12} - 10^{13})$~G after the collapse. An additional amplification of this primary magnetic field can occur at the SN explosion. Generally, this amplification is caused by a fast rotation of a supernova core~\cite{Bisnovatyi-Kogan:1971} that leads to a growth of the magnetic field up to $B \sim (10^{14} - 10^{15})$~G and a magnetorotational supernova explosion (see, e.g. Ref.~\cite{Bisnovatyi-Kogan:2018vvk,Obergaulinger:2018udn,Sawai:2015tsa}). However, a generation of such a strong magnetic field is also possible without the fast rotation of the supernova core~\cite{Obergaulinger:2014}.

The magnetic field can influence not only SN dynamics, but modify also neutrino processes which are allowed in a supernova matter. Early studies of the $\beta$-processes~\cite{Korovina:1964,FassioCanuto:1970wk,Matese:1969zz}, although done with certain 
simplifications, show that modifications of these processes can be significant. Later on, this was applied to various astrophysical objects. In particular, for the core-collapse supernova, a magnetic field influence on $\beta$-processes was studied 
in Refs.~\cite{Lai:1997mm,Roulet:1997sw,Lai:1998sz,Arras:1998mv,Goyal:1998nq,Gvozdev:1999md,Gvozdev:2002nu,Kotake:2004jt,Duan:2004nc,Duan:2005fc,Ognev:2016wlq}. In most studies, there were either numerical analyses or analytical calculations with simplified assumptions, for example, about a strength of the magnetic field. 
In the present paper we omit
restrictions on the magnetic field strength and consider any magnetic field strength discussed in applications to astrophysical objects. 
For such magnetic fields, we obtain simple analytical expressions for reaction rates of beta-processes as well as the energy and momentum transferred from neutrinos and antineutrinos to the matter. 
These analytical results allow us to analyze a magnetic-field influence on beta-processes for various models of a supernova explosion. Note that these expressions can be also applied to postmerger accretion discs. In previous studies, a magnetic-field impact was estimated based on individual $\beta$-processes. In a difference, we consider an effect of the magnetic field on the whole set of beta-processes~(\ref{eq:p1})--(\ref{eq:p4}) in this paper.

This paper is organized as follows. Simple analytical expressions for reaction rates of beta-processes~(\ref{eq:p1})--(\ref{eq:p4}), from which local number densities of nucleons (or a chemical composition of a matter) can be determined, heating rates by neutrinos and antineutrinos as well as momenta transferred from neutrinos and antineutrinos to the matter in dependence on the magnetic field strength
are collected in Sec.~\ref{sec:analytical-results}. In Sec.~\ref{sec:numerical-result}, we present numerical analysis of macroscopic quantities mentioned above under conditions of the core-collapse supernova with magnetic field strengths up to $B \sim 10^{16}$~G, reachable in strongly magnetized supernovae. These quantities depend on a vast amount of matter and neutrino parameters, many of which can be fixed by using the results of the 1D PROMETHEUS-VERTEX simulations~\cite{Hudepohl:2014}\footnote{https://wwwmpa.mpa-garching.mpg.de/ccsnarchive/archive.html}. This allows us to study the magnetic field influence on reaction rates, heating rates and momenta transferred to the matter. Note that these quantities are local and in the core-collapse supernova model considered, they depend on the distance from the proto-neutron star center and evolve in time after a bounce. The corresponding dependences are worked out numerically. We conclude in Sec.~\ref{sec:conclusions}. Properties of two basic functions entering analytical expressions are presented in Appendix~\ref{sec:properties-I-and-J}.  

\section{Analytical results}
\label{sec:analytical-results}

Macroscopic effects of neutrinos and antineutrinos on a supernova matter are described by a reaction rate~$\varGamma$ (the number of processes occurring in a unit volume per unit time) as well as 
the energy and momentum~$\mathcal{P}^\mu$ transferred from neutrinos or antineutrinos to a unit volume of the medium per unit time. These quantities for $\beta$-processes~(\ref{eq:p1})-(\ref{eq:p4}) can be expressed in terms of neutrino $\mathcal{K}_\nu$ and antineutrino $\bar{\mathcal{K}}_\nu$ medium emissivities~\cite{Ognev:2016wlq}:
\begin{equation}
\begin{gathered}
\label{Q1}
\varGamma^{(1)}
=  
\int  
\big[ 1 - f_\nu \big]  \, \mathcal{K}_\nu \, \frac{d^3 q}{(2 \pi)^{3}},
\\
\mathcal{P}^{(1)\,\mu}
= -  
\int  
q^\mu \big[ 1 - f_\nu \big]  \, \mathcal{K}_\nu \, \frac{d^3 q}{(2 \pi)^{3}} ,
\end{gathered}
\end{equation}
\begin{equation}
\begin{gathered}
\label{Q2}
\varGamma^{(2)}
= 
\int   
e^{(\omega - \delta\mu) / T} \, f_\nu \, \mathcal{K}_\nu \, \frac{d^3 q}{(2 \pi)^{3}} ,
\\
\mathcal{P}^{(2) \mu}
= 
\int  
q^\mu \, e^{(\omega - \delta\mu) / T} \, f_\nu \, \mathcal{K}_\nu \, \frac{d^3 q}{(2 \pi)^{3}} \,,
\end{gathered}
\end{equation}
\begin{equation}
\begin{gathered}
\label{Q3}
\varGamma^{(3)}
=  
\int  
\big[ 1 - \bar{f}_\nu \big]  \, \bar{\mathcal{K}}_\nu \, \frac{d^3 \bar{q}}{(2 \pi)^{3}} ,
\\
\mathcal{P}^{(3)\,\mu}
= -  
\int   
\bar{q}^\mu \big[ 1 - \bar{f}_\nu \big]  \, \bar{\mathcal{K}}_\nu \, \frac{d^3 \bar{q}}{(2 \pi)^{3}} ,
\end{gathered}
\end{equation}
\begin{equation}
\begin{gathered}
\label{Q4}
\varGamma^{(4)}
= 
\int  
e^{(\bar{\omega} + \delta\mu) / T} \, \bar{f}_\nu \, \bar{\mathcal{K}}_\nu \, \frac{d^3 \bar{q}}{(2 \pi)^{3}},
\\
\mathcal{P}^{(4) \mu}
= 
\int   
\bar{q}^\mu \, e^{(\bar{\omega} + \delta\mu) / T} \, \bar{f}_\nu \, \bar{\mathcal{K}}_\nu \, \frac{d^3 \bar{q}}{(2 \pi)^{3}} \,.
\end{gathered}
\end{equation}
Here, $q^{\mu} = (\omega, {\bf q})$ and $\bar{q}^{\mu} = (\bar{\omega}, \bar{{\bf q}})$ are the neutrino and antineutrino 4-momenta, $\delta\mu = \mu_e + \mu_p - \mu_n$, $\mu_e$, $\mu_p$ and $\mu_n$ are the chemical potentials of electrons, protons and neutrons, respectively, $T$ is the matter temperature, $f_\nu$ and $\bar{f}_\nu$ are the neutrino and antineutrino distribution functions. 
Throughout the paper, we use the system of units in which $\hbar = c = k_{\rm B} = 1$. Hereafter, parameters with the bar correspond to antiparticles~---~antineutrinos and positrons. Upper indices in $\varGamma^{(i)}$ and $\mathcal{P}^{(i) \mu}$ indicate the $\beta$-process from the set~(\ref{eq:p1})-(\ref{eq:p4}).

\begin{figure}[tb]
\centerline{
\includegraphics[width=0.85\columnwidth]
{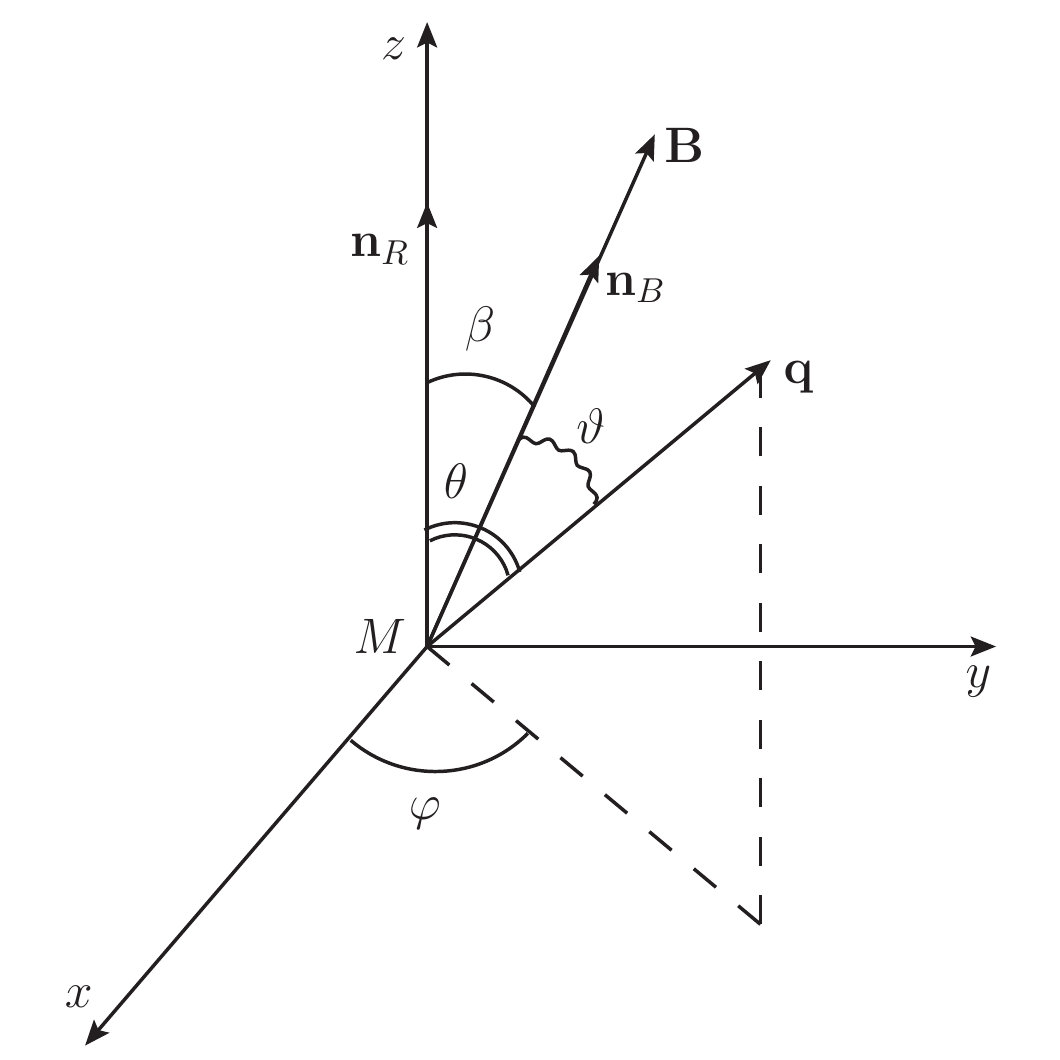} 
}
\caption{
Definitions of vectors and angles used in the analysis. 
Here, ${\bf n}_B = {\bf B}/B$ is the unit vector along the magnetic-field strength, ${\bf n}_R$ specifies the star radial direction, ${\bf q}$ is the neutrino momentum. 
The angles~$\beta$, $\theta$ and~$\vartheta$ between the vectors and the polar angle~$\varphi$ are also shown. 
}
\label{fig:1}
\end{figure}

The emissivities for neutrinos, $\mathcal{K}_\nu$, and antineutrinos, $\overline{\mathcal{K}}_\nu$, were obtained in Ref.~\cite{Ognev:2016wlq} for the case of non-degenerate protons and moderately degenerate electron-positron plasma. We consider an ultra-relativistic electron-positron plasma, therefore,
neutrino and antineutrino emissivities can be written as follows: 
\begin{equation}
\begin{gathered}
\mathcal{K}_{\nu} = G^2 N_p m_e^2 \, f_e(\omega) \, \varPhi(\omega/m_e, b, \vartheta) ,
\\
\overline{\mathcal{K}}_\nu = G^2 N_n m_e^2 \, \bar f_e(\bar\omega) 
\, \varPhi(\bar\omega/m_e, b, \bar\vartheta) \,.
\end{gathered}
\label{eq:emis-mod}
\end{equation}
In addition, we assume that neutrons are non-degenerate too and, hence, the number densities $N_n$ and $N_p$ of nucleons in Eq.~(\ref{eq:emis-mod}) are connected as:  
\begin{equation}
\label{Nn-Np-relation}
N_n = N_p \exp[(\mu_e - \delta\mu)/T]\,.
\end{equation}
Here, $m_e$ is the electron mass,
$f_e$ and $\bar f_e$ are the Fermi-Dirac distribution functions for electrons and positrons,
\begin{equation}
\begin{gathered}
\varPhi (x, b, \vartheta) = b \, \big[ \Theta(x - 1) - \Theta(x - \sqrt{1 + 2b}) \big]
\\
+ \, 2x^2 \, \Theta(x - \sqrt{1 + 2b}) - g_{va} b \cos\vartheta \, \Theta(x - 1) \,,
\end{gathered}
\end{equation}
where $\Theta(x)$ is the Heaviside step function, $b = B/B_e$ is the reduced magnetic field strength written in terms of critical Schwinger value, $B_e = m_e^2/e \simeq 4.41 \times 10^{13}$~G, 
$\vartheta (\bar{\vartheta})$ is the angle between the neutrino (antineutrino) momentum and the magnetic field direction, as shown in Fig.~\ref{fig:1}, $G^2 = G_F^2 \cos^2\theta_c\, (g_v^2 + 3 g_a^2) / (2 \pi)$, 
where $g_v$ and $g_a$ are the vector and axial constants entering the nucleon charged current, 
$\theta_c$ is the Cabibbo angle, $G_F$ is the Fermi constant, 
and $g_{va} = (g_a^2 - g_v^2) / (3 g_a^2 + g_v^2)$.

The neutrino~$f_\nu$ and antineutrino~$\bar{f}_\nu$ distribution functions introduced in Eqs.~(\ref{Q1})--(\ref{Q4}) are non-equilibrium. They are normalized on the local neutrino~$N_\nu$ and antineutrino~$\bar{N}_\nu$ number densities as follows:
\begin{equation}
N_\nu =
 \int f_\nu \, \frac{d^3 q}{(2 \pi)^{3}} \,,
\qquad 
\bar{N}_\nu =
\int \bar{f}_\nu \, \frac{d^3 \bar{q}}{(2 \pi)^{3}} .
\end{equation}
We assume that $f_\nu$ and $f_{\bar \nu}$ are spherically symmetric with the origin in the supernova center, neglecting a magnetic field and stellar rotation dependence on the neutrino propagation. 
It is useful to introduce angular moments: 
\begin{equation}
\label{eq:an-mom-1}
\chi_n \equiv \langle \chi^n \rangle = \frac{1}{N_\nu}
\int \chi^n \, f_\nu \, \frac{d^3 q}{(2 \pi)^{3}} ,
\end{equation}
\begin{equation}
\label{eq:an-mom-2}
\bar{\chi}_n \equiv \langle \bar{\chi}^n \rangle =  \frac{1}{\bar N_\nu} 
\int \bar{\chi}^n \, \bar{f}_\nu \, \frac{d^3 \bar{q}}{(2 \pi)^{3}} ,
\end{equation}
where $\chi = \cos\theta$, $\bar\chi = \cos\bar\theta$,~$\theta$ and~$\bar\theta$ are the angles between the neutrino and antineutrino momenta and radial direction, respectively (for neutrino see Fig.~\ref{fig:1}).

It is convenient also to introduce the energy moments:  
\begin{equation}
\label{eq:mom-en-def-1}
\omega_n \equiv \langle \omega^n \rangle = \frac{1}{N_\nu} 
\int \omega^n f_\nu \, \frac{d^3 q}{(2 \pi)^{3}} ,
\end{equation}
\begin{equation}
\label{eq:mom-en-def-2}
\bar{\omega}_n \equiv \langle \bar{\omega}^n \rangle =  \frac{1}{\bar{N}_\nu}
\int \bar{\omega}^n \, \bar{f}_\nu \, \frac{d^3 \bar{q}}{(2 \pi)^{3}}\,.
\end{equation}

As indicated in Ref.~\cite{Janka:1989}, the neutrino energy distribution can be approximated by a nominal Fermi-Dirac distribution:  
\begin{equation}
\label{eq:f_eta}
\omega^2 f_\nu \sim \frac{\omega^2}{1 + \exp ( \omega/T_\nu - \eta_\nu )} ,
\end{equation}
which is characterized by two effective parameters~$T_\nu$ and~$\eta_\nu$.
The alternative approximation, motivated by an analytic simplicity, for the neutrino energy spectrum was suggested in Ref.~\cite{Keil:2002in} and is called ``$\alpha$-fit'': 
\begin{equation}
\label{eq:f_nu}
\omega^2 f_\nu \sim 
(\omega / \omega_1)^{\alpha - 1} \, e^{- \alpha\, (\omega / \omega_1)} ,
\end{equation}
where $\omega_1 = \langle \omega \rangle$ is its averaged energy and $\alpha$ is the pinching parameter:  
\begin{equation}
\alpha = \frac{\omega_1^2}{\omega_2 - \omega_1^2}\,, 
\label{eq:pinch-par}
\end{equation}
depending on the second energy moment $\omega_2$~(\ref{eq:mom-en-def-1}). 
The notations~$\bar\alpha$, $\bar\omega_1$ and~$\bar\omega_2$ are used for antineutrinos.
In our analysis, we approximate the energy distributions of neutrinos and antineutrinos by ``$\alpha$-fit''. 

Under an assumption that the matter is in the local equilibrium, the energy distributions of ultra-relativistic electrons~$\varepsilon^2 f_e (\varepsilon)$ and positrons~$\bar\varepsilon^2 \bar f_e (\bar\varepsilon)$, where~$\varepsilon$ and~$\bar\varepsilon$ are the electron and positron energies, respectively, are described by Eq.~(\ref{eq:f_eta}), and they can be also approximated like in Eq.~(\ref{eq:f_nu}). 
As analysis in Ref.~\cite{Keil:2002in} shown, these two types of distributions,~(\ref{eq:f_eta}) and~(\ref{eq:f_nu}), are largely equivalent when neutrinos or antineutrinos are quasi-non-degenerate, but with a growth of the parameter~$\eta_\nu$ or~$\eta_{\bar\nu}$, respectively, a difference between the fits increases. For practical purposes of our calculations, the ``$\alpha$-fit''~(\ref{eq:f_nu}) introduced for the neutrino distribution can be safely adopted for electrons and positrons if the electron chemical potential satisfy the condition $\mu_e/T \lesssim 10$. Due to the matter electoneutrality, we have $\bar\mu_e/T \lesssim 0$ for positrons at any time and the distribution in the form of the ``$\alpha$-fit''~(\ref{eq:f_nu}) is also applied here. The averages of electron energy~$\varepsilon_1$ and its squared~$\varepsilon_2$ and the corresponding pinching parameter~$s$ entering the electron ``$\alpha$-fit''~(\ref{eq:f_nu}) are dependent on~$\mu_e$ and~$T$. Similar quantities~$\bar\varepsilon_1$, $\bar\varepsilon_2$, and~$\bar s$,  being dependent on~$\bar\mu_e$ and~$T$, are used for positrons.

It is convenient to define the ratios $\gamma = \varepsilon_1/\omega_1$ of the electron and neutrino averaged energies and $\bar\gamma = \bar\varepsilon_1/\bar\omega_1$ for positrons and antineutrinos   
as well as $\gamma_t = \varepsilon_1/T$ and $\bar\gamma_t = \bar\varepsilon_1/T$,  which are the electron and positron averaged energies in units of the temperature.

Despite electrons and positrons, assumed to be massless, are termalized in the magnetized medium, unmagnetized number densities of these particles:
\begin{equation}
N_0 = \frac{1}{\pi^{2}}
\int\limits_0^\infty f_e(\varepsilon) \, \varepsilon^2 d\varepsilon ,
\quad 
\bar{N}_0 = \frac{1}{\pi^{2}}
\int\limits_0^\infty \bar f_e (\bar\varepsilon) \, \bar\varepsilon^2 d\bar\varepsilon \,,
\end{equation}
are more convenient in applications.

\begin{figure}[tb]
\centerline{
\includegraphics[width=0.85\columnwidth]
{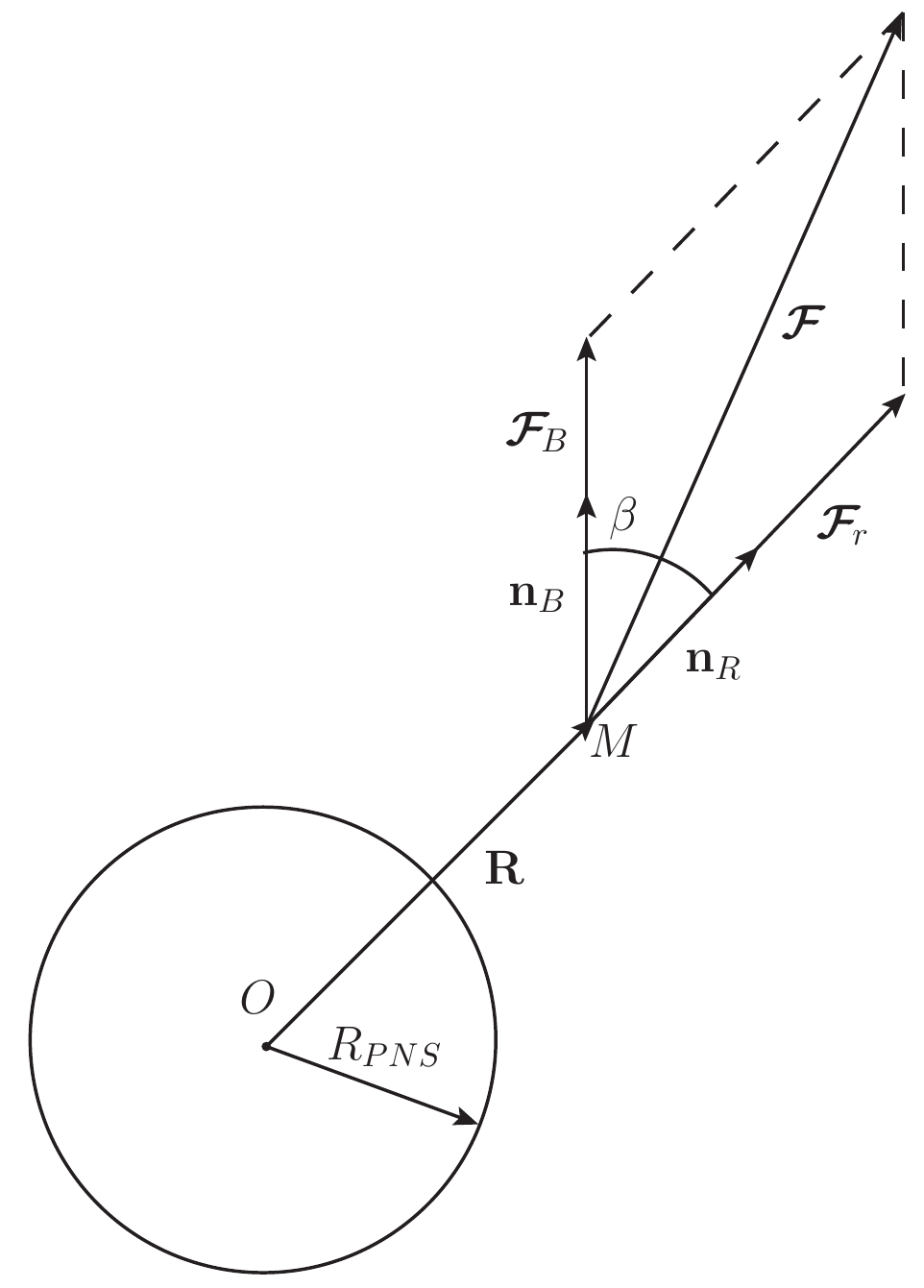} 
}
\caption{
The proto-neutron star (PNS) of radius~$R_{PNS}$ and momentum~$\boldsymbol{\mathcal{F}}$ transferred from neutrinos and antineutrinos to the matter at the point~$M$, being at the distance~$R$ from the PNS center.
${\bf n}_R$ and ${\bf n}_B$ are unit vectors in the radial direction from the PNS center and along the magnetic field strength, respectively.
}
\label{fig:2}
\end{figure}

Due to the symmetry assumed, calculations are carried out in the spherical coordinate system shown in Fig.~\ref{fig:1}, in which ${\bf n}_R$ is the unit vector in the star radial direction, ${\bf n}_B = {\bf B}/B$ is the unit vector along the magnetic field strength, and $({\bf n}_B {\bf n}_R) = \cos\beta$. 

For the reaction rates, we have: 
\begin{equation}\hspace{-12mm}
\begin{aligned}
\label{eq:g1}
\varGamma^{(1)} & = \ G^2 N_p N_0 \, \varepsilon_1^2 \, s^s \, \Gamma^{-1} (s)
\\
& \times  
\big[ I_{s-1, s} (\varepsilon_1, b)
- n_\nu \, I_{s + \alpha-4, s + \gamma\alpha} (\varepsilon_1, b)
\\
& +  g_{va} \cos\beta\, \chi_1\, n_\nu \,J_{s+\alpha-4, s+\gamma\alpha}(\varepsilon_1, b) \big] ,
\end{aligned}
\end{equation}
\vspace{-4mm}
\begin{equation}\hspace{-15mm}
\begin{aligned}
\label{eq:g2}
\varGamma^{(2)} & = \ G^2 N_n N_0 \, \varepsilon_1^2 \, e^{-\tau} \, s^s \, \Gamma^{-1}(s)
\\
&\times  \big[  n_\nu \, I_{s+\alpha-4, s+\gamma\alpha-\gamma_t}(\varepsilon_1, b)
\\
& - g_{va} \cos\beta\, \chi_1\, n_\nu \,J_{s+\alpha-4, s+\gamma\alpha-\gamma_t}(\varepsilon_1, b) \big] ,
\end{aligned}
\end{equation}
\vspace{-3mm}
\begin{equation}\hspace{-15mm}
\begin{aligned}
\label{eq:g3}
\varGamma^{(3)} & = \ G^2 N_n \bar{N}_0 \, \bar\varepsilon_1^2 \, \bar{s}^{\bar{s}} \,\Gamma^{-1}(\bar{s})
\\
& \times \big[ I_{\bar{s}-1, \bar{s}}(\bar\varepsilon, b) - \bar n_{\nu} 
\, I_{\bar{s}+\bar\alpha-4, \bar{s}+\bar{\gamma}\bar{\alpha}}(\bar\varepsilon, b)
\\
& + g_{va} \cos\beta\, \bar{\chi}_1 \, \bar n_{\nu} 
\, J_{\bar{s}+\bar\alpha-4, \bar{s}+\bar{\gamma}\bar{\alpha}}(\bar\varepsilon, b) \big] ,
\end{aligned}
\end{equation}
\vspace{-3mm}
\begin{equation}\hspace{-15mm}
\begin{aligned}
\label{eq:g4}
\varGamma^{(4)} & = \ G^2 N_p \bar{N}_0 \, \bar\varepsilon_1^2 \, e^{\tau} \, \bar{s}^{\bar{s}} 
\, \Gamma^{-1}(\bar{s})
\\
& \times \big[ \bar n_{\nu} 
\, I_{\bar{s}+\bar\alpha-4, \bar{s}+\bar{\gamma}\bar{\alpha}-\bar{\gamma}_t}(\bar\varepsilon, b)
\\
& - g_{va} \cos\beta\, \bar{\chi}_1\, \bar n_{\nu} 
\, J_{\bar{s}+\bar\alpha-4, \bar{s}+\bar{\gamma}\bar{\alpha}-\bar{\gamma}_t}(\bar\varepsilon, b) \big] ,
\end{aligned}
\end{equation}
and the energies $Q^{(i)} = \mathcal{P}^{(i)\,0}$ transferred from neutrino and antineutrino to the matter are as follows: 
\begin{equation}\hspace{-15mm}
\begin{aligned}
\label{eq:q1}
Q^{(1)} & = \ G^2 N_p N_0 \, \varepsilon_1^3 \, s^s \, \Gamma^{-1}(s)
\\
& \times  \big[ n_\nu \, I_{s+\alpha-3, s+\gamma\alpha}(\varepsilon_1, b)
- I_{s,s}(\varepsilon_1, b)
\\
& - g_{va} \cos\beta\, \chi_1\,n_\nu \, J_{s+\alpha-3, s+\gamma\alpha}(\varepsilon_1, b) \big] ,
\end{aligned}
\end{equation}
\vspace{-3mm}
\begin{equation}\hspace{-15mm}
\begin{aligned}
\label{eq:q2}
Q^{(2)} & = \ G^2 N_n N_0 \, \varepsilon_1^3 \, e^{-\tau} \, s^s \, \Gamma^{-1}(s)
\\
& \times \big[ n_\nu \, I_{s+\alpha-3, s+\gamma\alpha-\gamma_t}(\varepsilon_1, b)
\\
& - g_{va} \cos\beta\, \chi_1 \, n_\nu \, J_{s+\alpha-3, s+\gamma\alpha-\gamma_t}(\varepsilon_1, b) \big] ,
\end{aligned}
\end{equation}
\vspace{-3mm}
\begin{equation}\hspace{-15mm}
\begin{aligned}
\label{eq:q3}
Q^{(3)} & = \ G^2 N_n \bar{N}_0 \, \bar\varepsilon_1^3 \, \bar{s}^{\bar{s}} \,\Gamma^{-1}(\bar{s})
\\
& \times \big[ \bar n_{\nu} \, 
I_{\bar{s}+\bar\alpha-3, \bar{s}+\bar{\gamma}\bar{\alpha}}(\bar\varepsilon_1, b)
- I_{\bar{s}, \bar{s}}(\bar\varepsilon_1, b)
\\
& - g_{va} \cos\beta\, \bar{\chi}_1 \, \bar n_{\nu} \, 
J_{\bar{s}+\bar\alpha-3, \bar{s}+\bar{\gamma}\bar{\alpha}}(\bar\varepsilon_1, b) \big] ,
\end{aligned}
\end{equation}
\vspace{-3mm}
\begin{equation}\hspace{-15mm}
\begin{aligned}
\label{eq:q4}
Q^{(4)} & = \ G^2 N_p \bar{N}_0 \, \bar\varepsilon_1^3 \, e^{\tau} \, \bar{s}^{\bar{s}} 
\, \Gamma^{-1}(\bar{s})
\\
& \times \big[ \bar n_{\nu} \, 
I_{\bar{s}+\bar\alpha-3, \bar{s}+\bar{\gamma}\bar{\alpha}-\bar\gamma_t}(\bar\varepsilon_1, b)
\\
& - g_{va} \cos\beta\, \bar{\chi}_1 \, \bar n_{\nu} \, 
J_{\bar{s}+\bar\alpha-3, \bar{s}+\bar{\gamma}\bar{\alpha}-\bar\gamma_t}(\bar\varepsilon_1, b) \big] \,.
\end{aligned}
\end{equation}
Here, $\tau = \mu_e / T$, and the dimensionless neutrino and antineutrino number densities are introduced: 
\begin{equation}
n_\nu = 2\pi^2 \frac{\ (\alpha\gamma)^\alpha}{\Gamma(\alpha)}\, \frac{N_\nu}{\varepsilon_1^3} ,
\quad 
\bar n_\nu = 2\pi^2 \frac{\ (\bar\alpha \bar\gamma)^{\bar\alpha}}{\Gamma(\bar\alpha)}\, \frac{\bar N_\nu}{\bar\varepsilon_1^3} .
\end{equation}

The momentum $\boldsymbol{\mathcal{F}}^{(i)} = (\mathcal{P}^{(i)\,1}, \mathcal{P}^{(i)\,2}, \mathcal{P}^{(i)\,3})$ transferred from neutrinos and antineutrinos to the matter can be decomposed into two components, $\boldsymbol{\mathcal{F}}^{(i)} = \mathcal{F}_{B}^{(i)}\, {\bf n}_B + \mathcal{F}_{r}^{(i)}\, {\bf n}_R$, as shown in Fig.~\ref{fig:2}. 

Momenta transferred along the magnetic-field strength vector, $\mathcal{F}_{B}^{(i)}$, can be presented in the form: 
\begin{equation}
\begin{aligned}
\label{eq:fb1}
\mathcal{F}_{B}^{(1)} & = \ (g_{va}/3) \, G^2 N_p N_0 \, \varepsilon_1^3 \, s^s \, \Gamma^{-1}(s)
\, \big[ J_{s,s}(\varepsilon_1, b)
\\
& + (3/2) \, (\chi_2 - 1) \, n_\nu \, J_{s+\alpha-3, s+\gamma\alpha}(\varepsilon_1, b) \big] ,
\end{aligned}
\end{equation}
\vspace{-3mm}
\begin{equation}\hspace{-17mm}
\begin{aligned}
\label{eq:fb2}
\mathcal{F}_{B}^{(2)} & = (g_{va}/2) \, G^2 N_n N_0 \, \varepsilon_1^3 \, e^{-\tau} \, 
s^s \, \Gamma^{-1}(s)
\\
& \times (\chi_2 - 1) \, n_\nu \, J_{s+\alpha-3, s+\gamma\alpha-\gamma_t}(\varepsilon_1, b) ,
\end{aligned}
\end{equation}
\vspace{-3mm}
\begin{equation}
\begin{aligned}
\label{eq:fb3}
\mathcal{F}_{B}^{(3)} & = \ (g_{va}/3) \, G^2 N_n \bar{N}_0 \, \bar\varepsilon_1^3 \, \bar{s}^{\bar{s}}
\,\Gamma^{-1}(\bar{s}) \, \big[ J_{\bar{s}, \bar{s}}(\bar\varepsilon_1, b)
\\
& + (3/2) \, (\bar{\chi}_2 - 1 ) \, \bar{n}_{\nu} 
\, J_{\bar{s}+\bar\alpha-3, \bar{s}+\bar{\gamma}\bar{\alpha}}(\bar\varepsilon_1, b) \big] ,
\end{aligned}
\end{equation}
\vspace{-3mm}
\begin{equation}\hspace{-17mm}
\begin{aligned}
\label{eq:fb4}
\mathcal{F}_{B}^{(4)} & = (g_{va}/2) \, G^2 N_p \bar{N}_0 \, \bar{\varepsilon}_1^3 \, e^{\tau} 
\, \bar{s}^{\bar{s}} \, \Gamma^{-1}(\bar{s})
\\
& \times (\bar{\chi}_2 - 1) \, \bar{n}_{\nu} 
\, J_{\bar{s}+\bar\alpha-3, \bar{s}+\bar{\gamma}\bar{\alpha}-\bar\gamma_t}(\bar\varepsilon_1, b) .
\end{aligned}
\end{equation}
In the radial direction,~$\mathcal{F}_{r}^{(i)}$ are as follows: 

\begin{eqnarray}
\hspace{-3mm}
\label{eq:fr1}
\mathcal{F}_{r}^{(1)} & = & \ G^2 N_p N_0 \, n_\nu \, \varepsilon_1^3 \, s^s \, \Gamma^{-1}(s) 
\\ \nonumber
& \times & \big[ (g_{va} / 2) \cos\beta (1 - 3 \chi_2) \, J_{s+\alpha-3, s+\gamma\alpha}(\varepsilon_1, b) \\ \nonumber
& + & \chi_1 \, I_{s+\alpha-3, s+\gamma\alpha}(\varepsilon_1, b) \big] , 
\end{eqnarray}
\vspace{-3mm}
\begin{eqnarray}
\label{eq:fr2}
\mathcal{F}_{r}^{(2)} & = & \ G^2 N_n N_0 \, n_\nu \, \varepsilon_1^3 \, e^{-\tau} \, s^s \, \Gamma^{-1}(s)
\\ \nonumber
& \times & \big[ (g_{va} / 2) \cos\beta (1 - 3 \chi_2) \, 
J_{s+\alpha-3, s+\gamma\alpha-\gamma_t}(\varepsilon_1, b)
\\ \nonumber
& + & \chi_1 \, I_{s+\alpha-3, s+\gamma\alpha-\gamma_t}(\varepsilon_1, b) \big] ,
\end{eqnarray}
\vspace{-3mm}
\begin{eqnarray}\hspace{-3mm}
\label{eq:fr3}
\mathcal{F}_{r}^{(3)} & = & \ G^2 N_n \bar{N}_0 \, \bar{n}_{\nu} \, \bar\varepsilon_1^3 
\, \bar{s}^{\bar{s}} \,\Gamma^{-1}(\bar{s})
\\ \nonumber
& \times & \big[ (g_{va} / 2)\cos\beta(1 - 3 \bar{\chi}_2) 
\, J_{\bar{s}+\bar\alpha-3, \bar{s}+\bar{\gamma}\bar{\alpha}}(\bar\varepsilon_1, b)
\\ \nonumber
& + & \bar{\chi}_1 \, 
I_{\bar{s}+\bar\alpha-3, \bar{s}+\bar{\gamma}\bar{\alpha}}(\bar\varepsilon_1, b) \big] ,
\end{eqnarray}
\vspace{-3mm}
\begin{eqnarray}
\label{eq:fr4}
\mathcal{F}_{r}^{(4)} & = &\ G^2 N_p \bar{N}_0 \, \bar{n}_{\nu} \, \bar{\varepsilon}_1^3 
\, e^{\tau} \, \bar{s}^{\bar{s}} \, \Gamma^{-1}(\bar{s})
\\ \nonumber
& \times & \big[ (g_{va} / 2) \cos\beta (1 - 3 \bar{\chi}_2) 
\, J_{\bar{s}+\bar\alpha-3, \bar{s}+\bar{\gamma}\bar{\alpha}-\bar\gamma_t}(\bar\varepsilon_1, b)
\\ \nonumber
& + & \bar{\chi}_1 
\, I_{\bar{s}+\bar\alpha-3, \bar{s}+\bar{\gamma}\bar{\alpha}-\bar\gamma_t}(\bar\varepsilon_1, b) \big] .
\end{eqnarray}

The magnetic-field dependence enters the rates, energies and momenta above through the functions: 
\begin{gather}
\begin{aligned}
\label{eq:i11}
& I_{k,\varkappa} (\varepsilon_1, b) = \varkappa^{-k-3}\, \Gamma (k+3,\varkappa\, z_b) 
\\
& \hspace{11mm}
+ \varkappa^{-k-1}\, \frac{b\,m_e^2}{2\varepsilon_1^2} 
\Big[ \Gamma(k+1) - \Gamma(k+1,\varkappa\, z_b) \Big]  ,
\end{aligned}
\\
\label{eq:i21}
\hspace{-30mm}
J_{k,\varkappa}(\varepsilon_1, b) = \varkappa^{-k-1}\, \frac{\,b\,m_e^2}{2\varepsilon_1^2} \, \Gamma(k+1) ,
\end{gather}
where $z_b = (m_e/\varepsilon_1) \sqrt{1+2b}$ and $\Gamma(x, y)$ is the incomplete Gamma-function~\cite{Abramowitz:1970}, $\Gamma(x) = \Gamma(x, 0)$. The functions~(\ref{eq:i11}) and~(\ref{eq:i21}) are discussed in details in Appendix~\ref{sec:properties-I-and-J}. 
In particular, the functions $I_{k,\varkappa} (\varepsilon_1, b)$ and $J_{k,\varkappa}(\varepsilon_1, b)$  are significantly modified by the magnetic field at $B \gg B_e$. The dependence of these functions on the field strength is defined through the dimensionless parameters:
\begin{eqnarray}
\eta = \varkappa \left ( m_e/\varepsilon_1 \right ) \sqrt{2b} \,, \qquad
\bar\eta = \bar\varkappa \left ( m_e/\bar\varepsilon_1 \right ) \sqrt{2b} \,, 
\label{eq:par-m-f}
\end{eqnarray}
for neutrinos and antineutrinos, respectively. As follows from these definitions, 
these parameters increase with a growth of the magnetic field strength and the degeneracy of leptons while the increase of average energy of electron-positron plasma reduces their values. We are unable explicitly predict the values of $\eta$ and $\bar{\eta}$ as these quantities, in addition to the field strength, are implicitly dependent on the distance from the PNS center and evolve in time. A combine effect is difficult to determine without numerical estimations. 

It should be noted that Eqs.~(\ref{eq:g1})--(\ref{eq:fr4}) are valid for the magnetic field of an arbitrary, physically motivated strength. It turns out that a dependence on the magnetic field and its spatial configuration is sufficiently simple. Indeed, due to the symmetry of the problem, only the reduced magnetic field strength, $b = B/B_e$, and its relative direction, through $\cos\beta$, enter Eqs.~(\ref{eq:g1})--(\ref{eq:fr4}). 

To have a matter transparent for neutrinos, it is necessary to put $n_\nu = \bar n_\nu = 0$.
In this case, all the radial components $\mathcal{F}_r^{(i)}$ of the momentum transferred vanish.  
The momenta $\mathcal{F}_B^{(i)}$ along the magnetic-field strength vector, reaction rates~$\varGamma^{(i)}$ and energies~$Q^{(i)}$ for the processes with the neutrino emission coincide with ones in Ref.~\cite{Ognev:2016wlq}, when the approximation~(\ref{eq:f_nu}) for the distribution functions of electrons and positrons is not used.

\section{Numerical results}
\label{sec:numerical-result}

It is easy to see that analytical results~(\ref{eq:g1})-(\ref{eq:fr4}) depend on different parameters of a matter and neutrino radiation such as the electron chemical potential~$\mu_e$, temperature~$T$, unmagnetized number densities of protons~$N_p$ and neutrons~$N_n$, neutrino~$N_\nu$ and antineutrino~$\bar{N}_\nu$ number densities, the first two angular moments of neutrino~$\chi_{1,2}$ and antineutrino~$\bar{\chi}_{1,2}$, the first two energy moments of neutrino~$\omega_{1,2}$ and antineutrino~$\bar\omega_{1,2}$. It is possible to fix their values by using the data provided by numerical simulations. For this purpose, we use the results of the 1D PROMETHEUS-VERTEX simulations~\cite{Hudepohl:2014} in which a model for the 27~$M_{\odot}$ progenitor was emploied~\cite{Woosley:2002zz}. The final neutron star has the baryonic mass equal to 1.76~$M_{\odot}$. Note that data for the 1.76~$M_{\odot}$ neutron star are representative and reproduce the results of modeling for other masses of SN progenitors. The nuclear equation of state is taken from Ref.~\cite{Lattimer:1991nc} with compressibility modulus $K = 220$~MeV. As self-consistent spherically symmetric models do not explode except for a few exceptional cases of low-mass progenitors, the explosion has to be initiated at $t = 0.5$~s after the core-bounce in an artificial way.
These simulations do not take into account the influence of the magnetic field. Thus, we assume that the magnetic-field influence on both neutrino and matter parameters is insignificant.
Based on the data available, it is possible to reduce the existing number of parameters to two only: the distance~$R$ from the PNS center and the time~$t$ after a bounce. The other two parameters in our problem are the magnetic field strength $B$ and angle $\beta$ between the radius-vector and field direction.

As follows from simulations, assumptions about non-degeneracy of the nucleon matter and moderate degeneracy of electrons are valid outside the proto-neutron star, i.e. above a sphere of the radius $R \sim 16$~km. It should be noted that inside this sphere neutrinos are in the thermodynamic equilibrium with the SN matter, therefore, neutrinos do not affect a matter in this region because the processes of the neutrino absorption and emission compensate each other. It follows from the simulations~\cite{Hudepohl:2014} that the electron-positron plasma is no longer ultra-relativistic in the supernova outer part, at $R \gtrsim 600$~km. 
In this region a neutrino flux sufficiently decreases and the neutrino interaction with the matter becomes ineffective. Hence, Eqs.~(\ref{eq:g1})--(\ref{eq:fr4}) are valid in the supernova region where neutrinos can affect the matter and influence on the supernova dynamics. Combining together, we determine the spherical layer bounded by two homocentric spheres with the radii $R_1 = 16$~km and $R_2 = 600$~km, in which our analysis is performed.

\subsection{Gain radius}

\begin{figure}[tb]
\centerline{
\includegraphics[width=0.85\columnwidth]{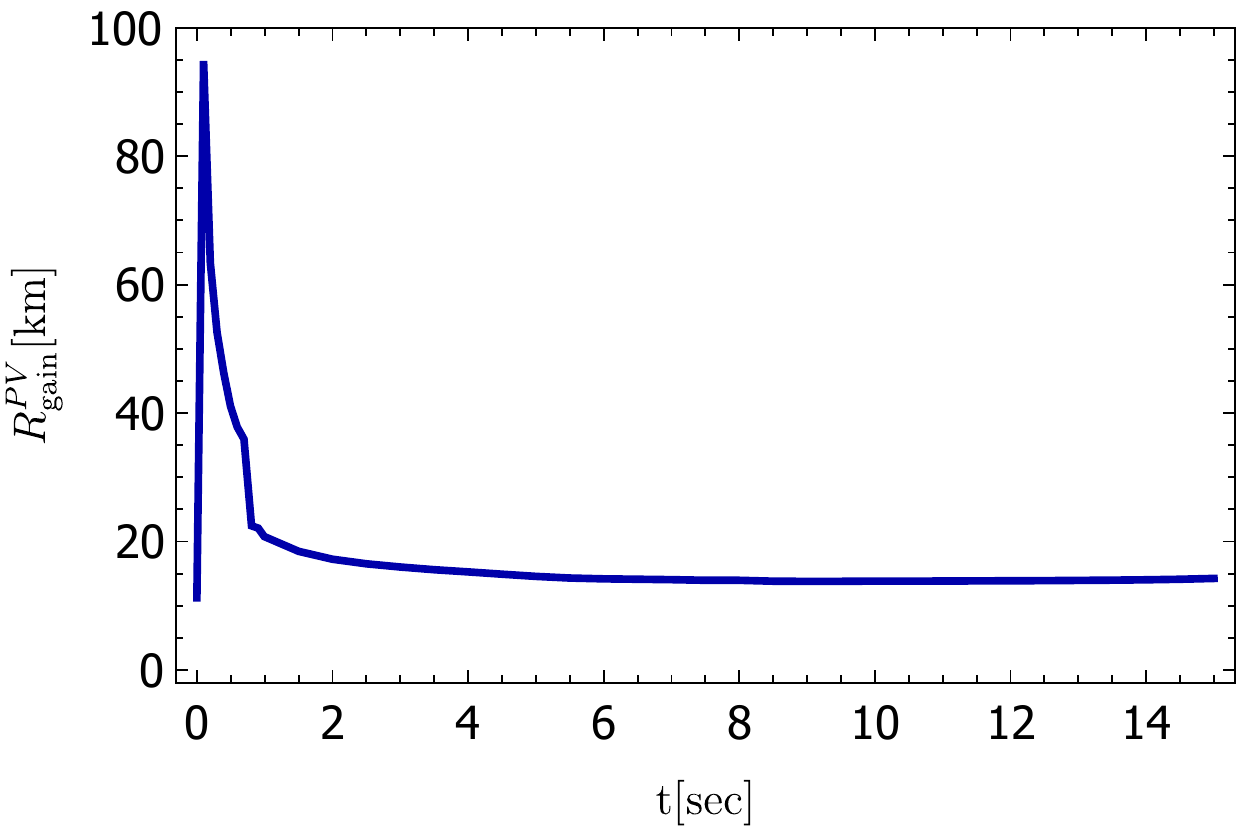} 
}
\caption{The gain radius $R_{\rm gain}^{PV}$ from PROMETHEUS-VERTEX simulations~\cite{Hudepohl:2014}.}
\label{fig:plot4}
\end{figure}

\begin{figure}[tb]
\centerline{
\includegraphics[width=0.85\columnwidth]{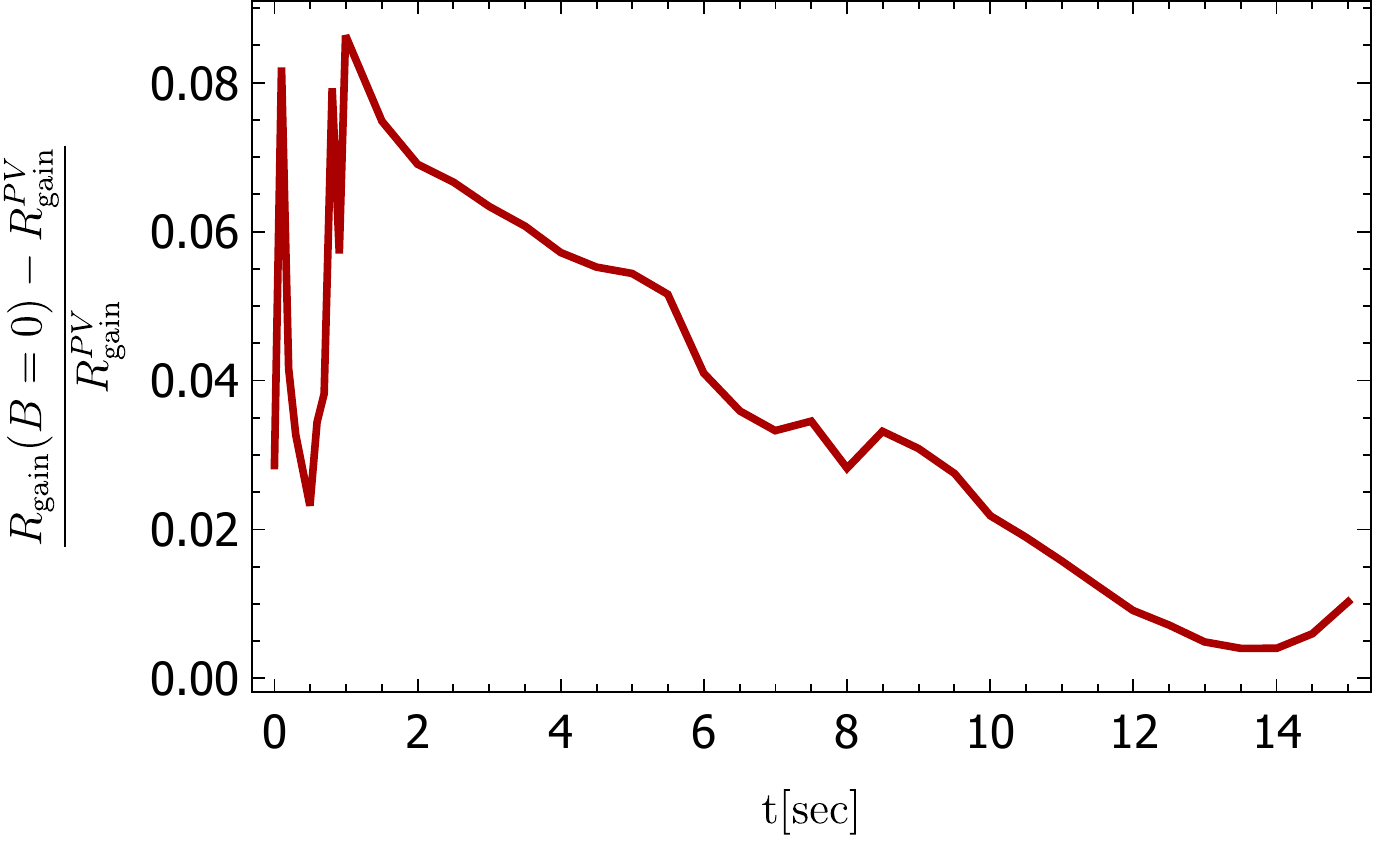} 
}
\caption{The deviation of a gain radius $R_{\rm gain} (B = 0)$ for a matter without the magnetic field, obtained by analytical calculations, from the gain radius $R_{\rm gain}^{PV}$ from PROMETHEUS-VERTEX simulations~\cite{Hudepohl:2014}.}
\label{fig:plot3}
\end{figure}

A spherical boundary between neutrino cooling and heating layers in supernova, where $Q^{(1)} + Q^{(2)} + Q^{(3)} + Q^{(4)} = 0$, has a radius~$R_{\rm gain}$, called the gain radius. 
The gain radius $R_{\rm gain}^{PV}$ from the PROMETHEUS-VERTEX simulations in dependence on the time after a bounce~\cite{Hudepohl:2014} is presented in Fig.~\ref{fig:plot4}. 
In Fig.~\ref{fig:plot3}, we compare a temporal evolution of $R_{\rm gain} (B = 0)$ for the matter without the field, obtained by analytical calculations, with the gain radius $R_{\rm gain}^{PV}$. The deviation of our results from the numerical ones is less than~8\%.
This difference is due to other, non-URCA, processes with neutrinos and antineutrinos included in the numerical code~\cite{Hudepohl:2014}. The gain radius obtained by us is from the requirement of vanishing the total energy of $\beta$-processes only. 
This analysis can be considered as a consistency check of the approach developed here and numerical simulations~\cite{Hudepohl:2014}. 
\begin{figure}[tb]
\centerline{
\includegraphics[width=0.85\columnwidth]{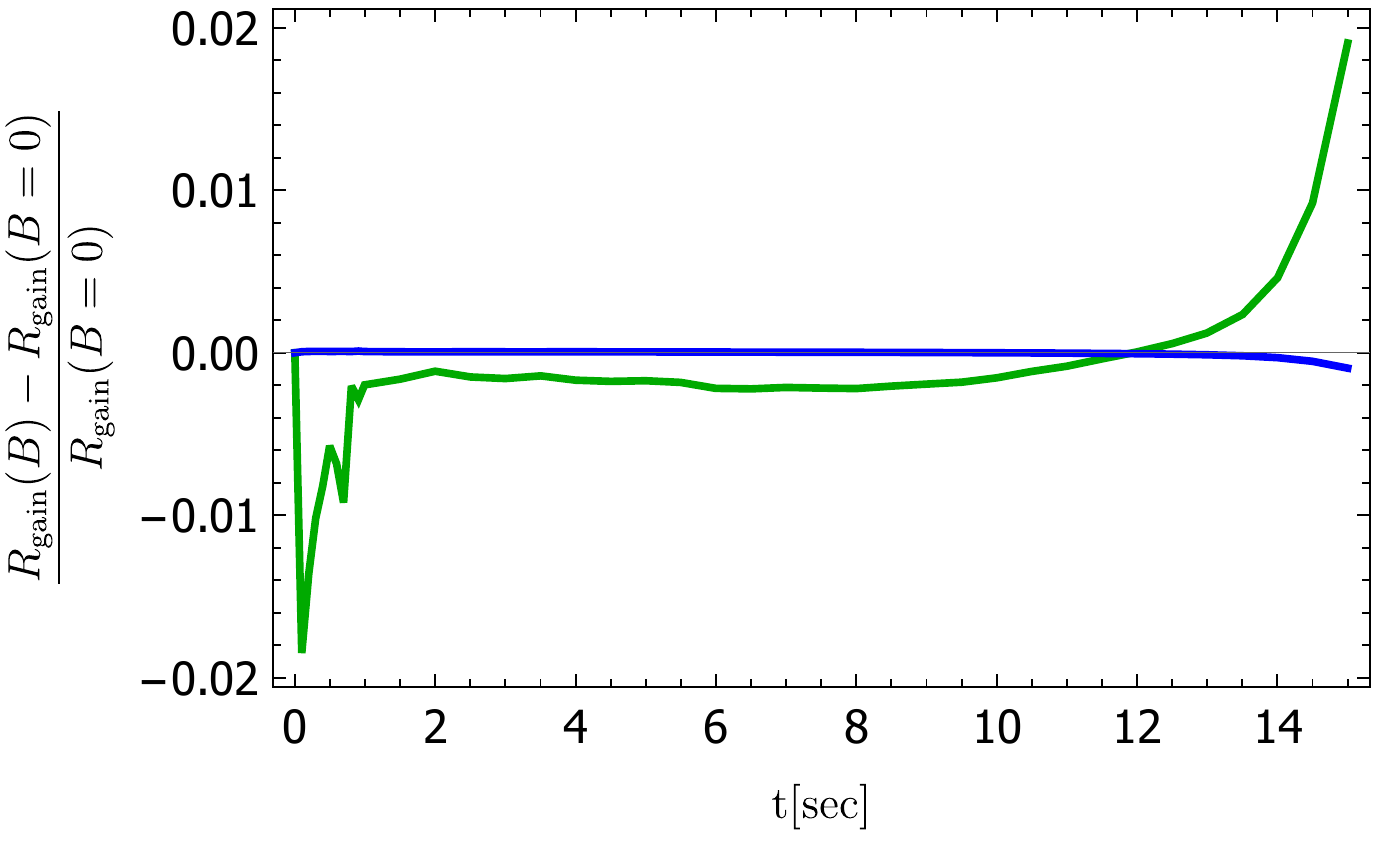} 
}
\caption{The deviation of a gain radius $R_{\rm gain} (B)$ in the presence of the magnetic field with $\cos\beta = 1$ from an unmagnetized one. Magnetic field strengths are $B = 10^{15}$~G (blue line) and $B = 10^{16}$~G (green line).}
\label{fig:plot5}
\end{figure}

An influence of the magnetic field on the gain radius, $R_{\rm gain}$, is shown in Fig.~\ref{fig:plot5}. To be definite, we use the magnetic-field configuration with $\cos\beta = 1$ in our analysis but the result obtained remains the same for other field directions as well. The maximal deviation from the unmagnetized case is limited by 2\% for the magnetic field strength $B = 10^{16}$~G. It should be noted that such a magnetic field strength is difficult to reach in supernova simulations. Thus, the analysis of a magnetic-field influence on $\beta$-processes below is performed for $B = 10^{15}$~G, which is more realistic for supernova conditions. 
Note that configurations of the magnetic field following from supernova simulations are quite different and strongly depend on many parameters. In calculations presented here, we restrict ourselves with a constant magnetic field, having a fixed direction to the star radius, i.\,e. the fixed angle~$\beta$.

\subsection{Total reaction rate and energy}

The reaction rates of beta-processes~(\ref{eq:g1})--(\ref{eq:g4}) determine the chemical composition of a supernova matter. As a quantity characterizing the influence of these processes on the matter, we consider the proton-to-neutron transition rate:
\begin{equation}
\varGamma(B, \cos\beta) = \varGamma_{p \to n} =
\varGamma^{(1)} - \varGamma^{(2)} - \varGamma^{(3)} + \varGamma^{(4)} .
\label{eq:sum-gamma}
\end{equation}
\begin{figure}[tb]
\centerline{
\includegraphics[width=0.85\columnwidth]{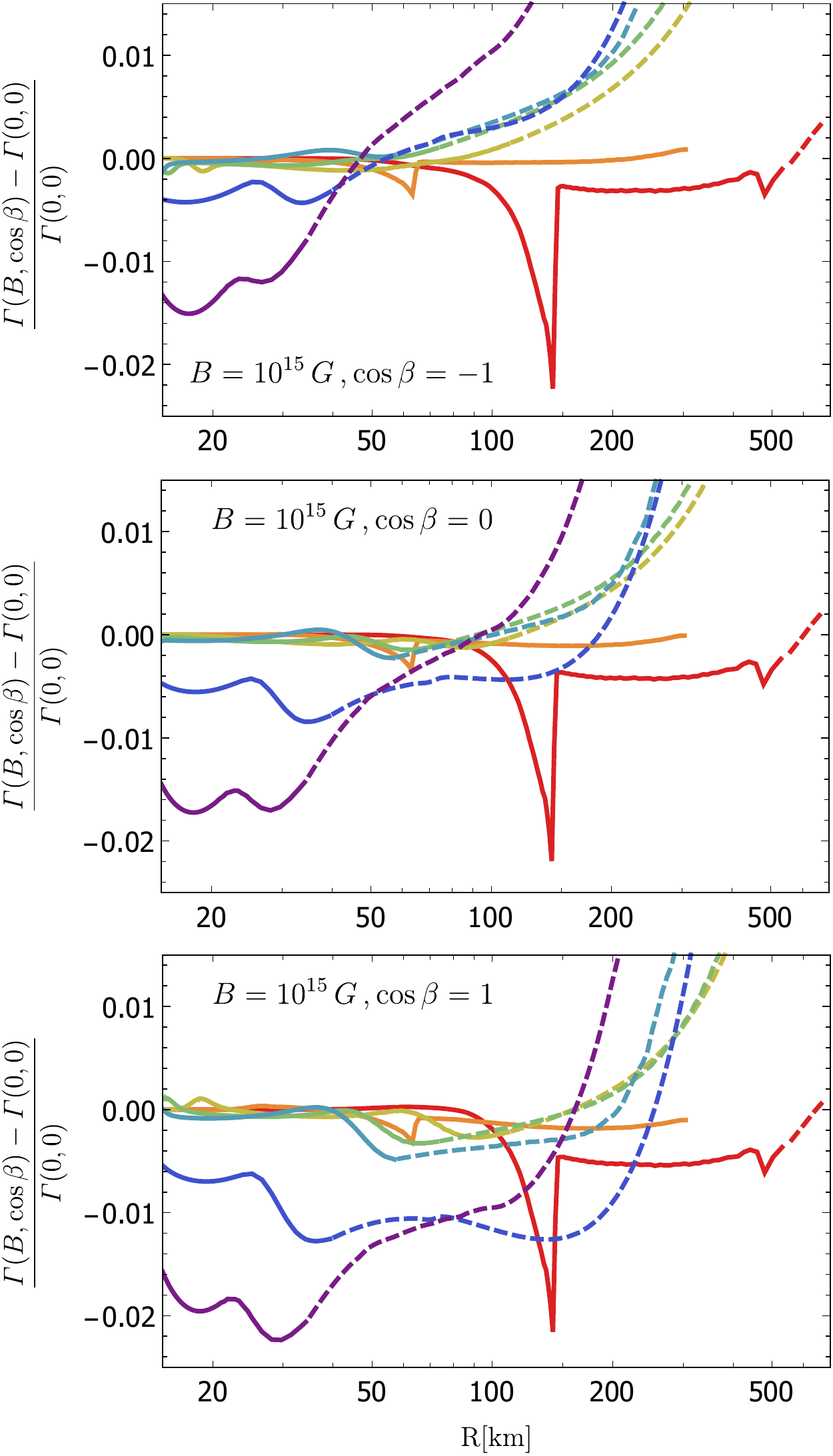} 
}
\caption{The relative deviation of the proton-to-neutron reaction rate of beta-processes in the magnetic field from the unmagnetized case as a function of distance~$R$ from the PNS center for several values of the time~$t$ after a bounce and different directions of the magnetic field. Configurations of magnetic field are $\cos\beta = -1$~(top panel), $\cos\beta = 0$ (middle panel) and $\cos\beta = 1$ (bottom panel). {\it Red lines:} $t = 0.1$~sec; {\it orange:} $t = 0.5$~sec; {\it yellow:} $t = 1.5$~sec; {\it green:} $t = 4$~sec; {\it cyan:} $t = 5.5$~sec; {\it blue:} $t = 10$~sec; {\it violet:} $t = 13$~sec.} 
\label{fig:3} 
\end{figure}

The relative deviations of $\varGamma (B, \cos\beta)$ from the reference value $\varGamma (0,0)$ in dependence on the distance~$R$ (in km) from the PNS center for several values of time after a bounce ($t = 0.1,\, 0.5,\, 1.5,\, 4.0,\, 5.5,\, 10,\, 13$~sec) and three directions of the magnetic field ($\cos\beta = -1, 0, 1$) are presented in Fig.~\ref{fig:3}. The dashed parts of the lines correspond to supernova regions where the electron-positron plasma is no longer ultrarelativistic. As seen in Fig.~\ref{fig:3}, the magnetic field affects significantly the reaction rates of the proton-to-neutron transition in dashed regions only, so one should perform corresponding calculations differently there. 

For solid lines, the influence of the magnetic field becomes more pronounced just after a bounce in region of the stalled shock wave and in a more later time, when the supernova has already cooled down through the neutrino emission.
Apparently, such a behavior is connected with the low-temperature region in the pre-shock environment
and at the end of the neutrino cooling phase. Nevertheless, even for these times, the modification of the proton-to-neutron transition rate by the magnetic field reaches a few percents only. Moreover, the magnetic field suppresses the neutron production in comparison with the unmagnetized case. It should be noted also that the direction of the magnetic field can also affect the proton-to-neutron transition.

\begin{figure}[tb]
\centerline{
\includegraphics[width=0.83\columnwidth]{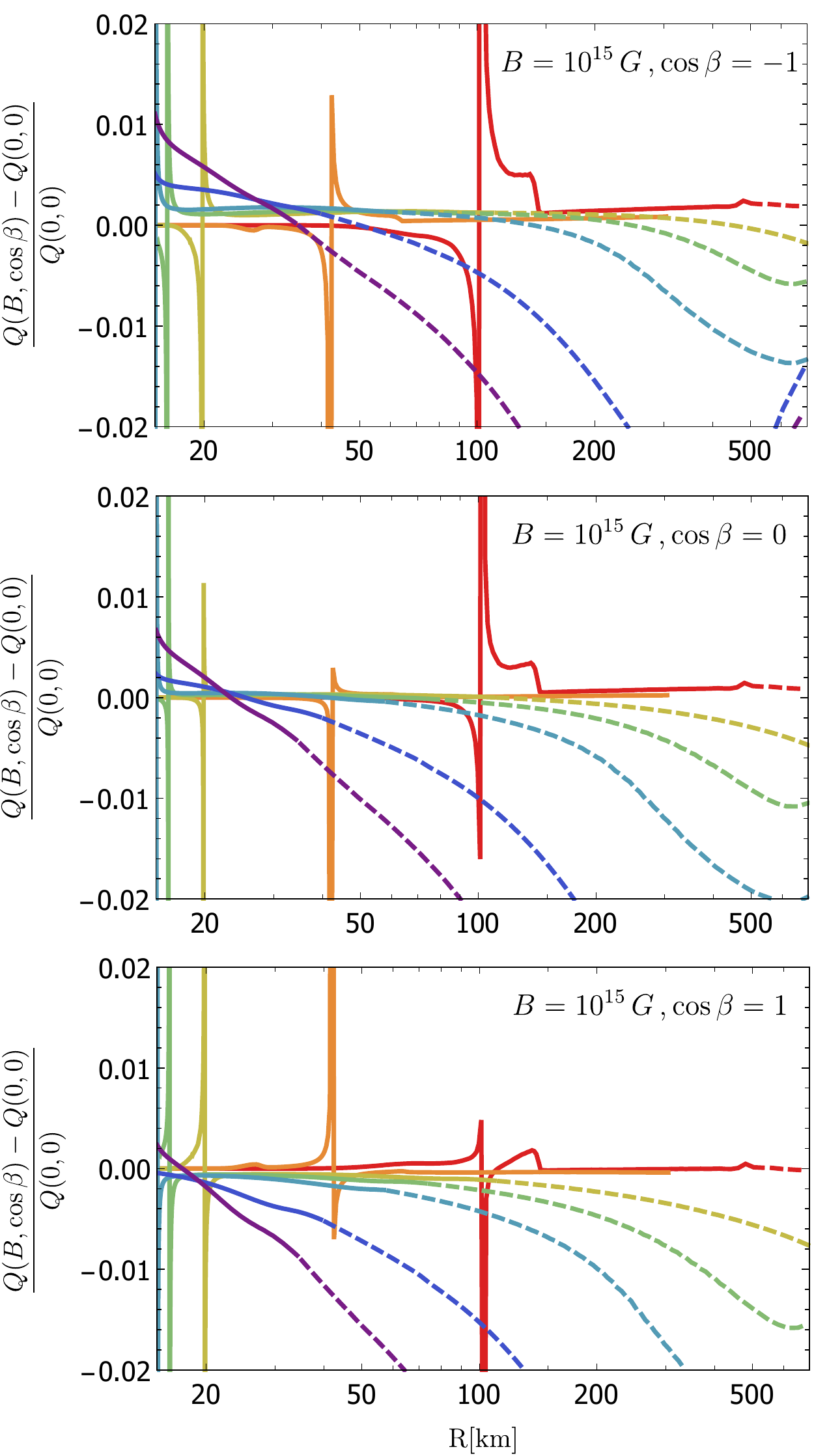} 
}
\caption{
The relative deviation of the total energy transferred from neutrinos and antineutrinos to the matter. 
The legend is the same as in Fig.~\ref{fig:3}.
} 
\label{fig:plot6} 
\end{figure}

The other important for supernova modeling quantity is the energy transferred from neutrinos and antineutrinos to the matter, known also as the heating rate. For the $\beta$-processes (\ref{eq:p1})--(\ref{eq:p4}), it has the form: 
\begin{equation}
Q (B, \cos\beta) = Q^{(1)} + Q^{(2)} + Q^{(3)} + Q^{(4)} .
\end{equation}
The relative deviations of the total energy, calculated in the presence of a magnetic field, from its representative value $Q (0, 0)$, corresponding to the unmagnetized matter, for several values of the time after a bounce ($t = 0.1$, 0.5, 1.5, 4.0, 5.5, 10, 13~sec) and three directions of the magnetic field ($\cos\beta = -1, 0, 1$) are shown in Fig.~\ref{fig:plot6}. The spikes are related with the gain radius where the total energy $Q (B, \cos\beta)$ goes through zero and changes its sign.

As seen in Figs.~\ref{fig:3} and~\ref{fig:plot6}, the character of the magnetic field influence on the matter heating has similarities with the reaction rates. Note that the magnetic field can increase the matter heating which is important for the explosion mechanism, but numerically this effect is small and, at most, reaches one percent only.

Some comments about the reaction~$\Gamma^{(i)}$ and heating~$Q^{(i)}$ rates of the individual $\beta$-processes~(\ref{eq:p1})--(\ref{eq:p4}) are in order. 
The magnetic-field influence on~$\Gamma^{(i)}$ and~$Q^{(i)}$ is consistent with the results obtained earlier (see, Refs.~\cite{Roulet:1997sw,Lai:1998sz,Duan:2004nc,Ognev:2016wlq}),  
in particular, these rates for each process separately are suppressed by the magnetic field. The calculations done give the suppression up to a few tens of percents at a maximum, which, nevertheless, is an order of magnitude larger than our estimates for the total rates of the beta-processes~(\ref{eq:p1})--(\ref{eq:p4}), presented in Figs.~\ref{fig:3} and~\ref{fig:plot6}. 
The further substantial reduction in the total rates is due to the different signs of the reaction and heating rates in the individual processes.  

\subsection{Total momentum}

The total momentum transferred from neutrinos and antineutrinos to the matter has the form: 
\begin{equation}
\boldsymbol{\mathcal{F}} = \boldsymbol{\mathcal{F}}^{(1)} + \boldsymbol{\mathcal{F}}^{(2)} + \boldsymbol{\mathcal{F}}^{(3)} + \boldsymbol{\mathcal{F}}^{(4)}\,.
\end{equation}
Each vector can be decomposed as follows (see Fig.~{\ref{fig:2}}):   
\begin{equation}
\boldsymbol{\mathcal{F}}^{(i)} = \mathcal{F}_B^{(i)}\, {\bf n}_B + \mathcal{F}_r^{(i)}\, {\bf n}_R\,.
\end{equation}
The radial component, $\mathcal{F}_r^{(i)}$, is determined by the radial asymmetry of the neutrino and antineutrino distribution functions, therefore, it vanishes in the matter transparent for  neutrinos and antineutrinos. 
The component $\mathcal{F}_B^{(i)}$ along the magnetic field direction is originated by the asymmetry of the neutrino emission and absorption caused by the magnetic field, and it is absent in the unmagnetized matter. 
\begin{figure}[tb]
\centerline{
\includegraphics[width=0.85\columnwidth]
{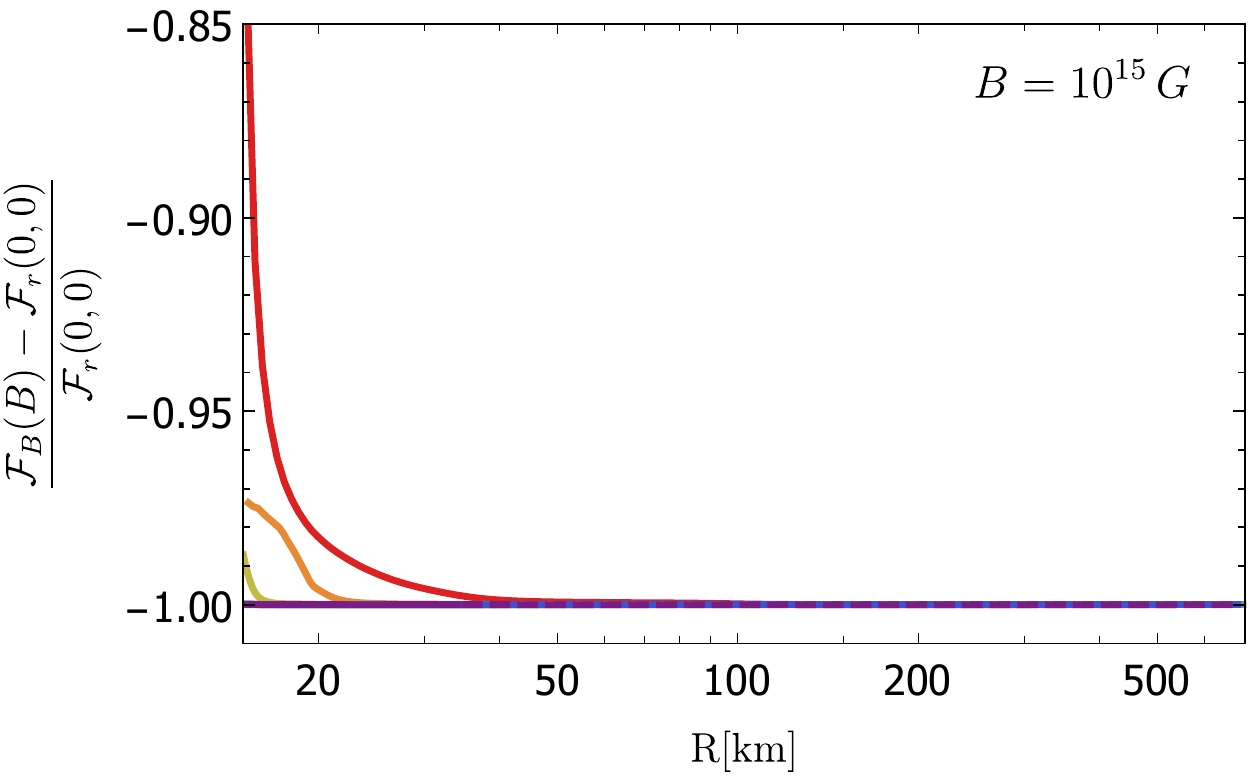} 
}
\caption{
The total momentum projection along the field, $\mathcal{F}_B (B)$, compared with the momentum projection $\mathcal{F}_r (0, 0)$ on the radial direction. 
The legend is the same as in Fig.~\ref{fig:3}.}
\label{fig:plot7}
\end{figure}
As shown in Fig.~\ref{fig:plot7}, $\mathcal{F}_r$ dominates over $\mathcal{F}_B$ in the supernova matter. The most sizable effect of~$\mathcal{F}_B$ is in the first few seconds after a bounce in the vicinity of a proto-neutron star surface.
Despite $\mathcal{F}_B$ is relatively small, it can modify the supernova dynamics, namely, it causes an additional non-radial motion of the supernova matter. In particular, for a toroidal configuration of the magnetic field,~$\mathcal{F}_B$ contributes into a supernova rotation. 
\begin{figure}[tb]
\centerline{
\includegraphics[width=0.85\columnwidth]
{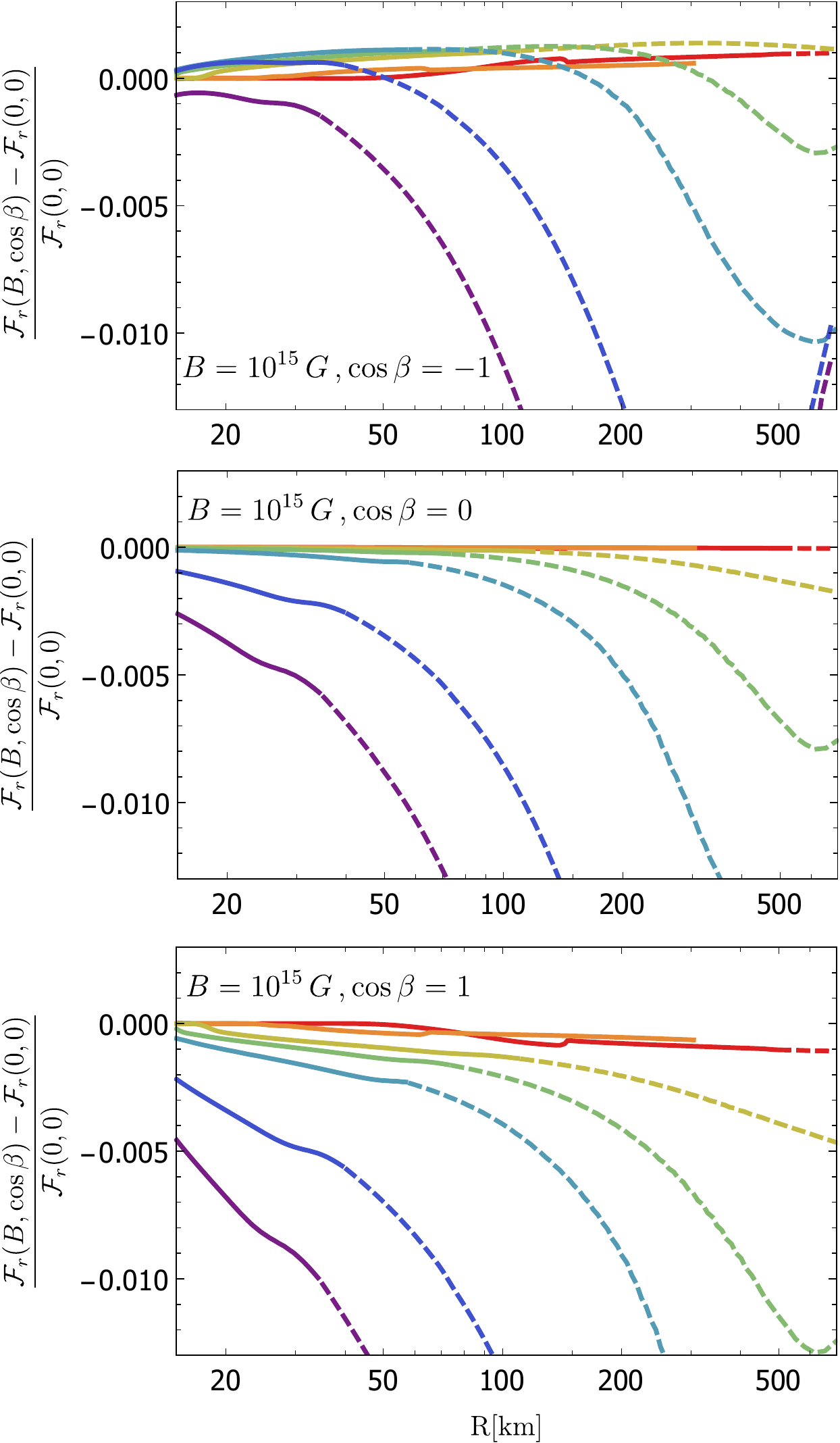} 
}
\caption{
The relative deviation of the radial component, $\mathcal{F}_r (B, \cos\beta)$, of the total momentum transferred from neutrinos and antineutrinos to the matter 
from its representative value $\mathcal{F}_r (0, 0)$, corresponding to the unmagnetized matter. 
The legend is the same as in Fig.~\ref{fig:3}.
} 
\label{fig:plot8} 
\end{figure}
In Fig.~\ref{fig:plot8}, we present the relative deviation of the total radial momentum, $\mathcal{F}_r (B, \cos\beta)$, transferred from neutrinos and antineutrinos to the matter in the presence of the magnetic field, from its representative value $\mathcal{F}_r (0, 0)$, corresponding to the unmagnetized matter, 
for several values of the time after a bounce ($t = 0.1$, 0.5, 1.5, 4.0, 5.5, 10, 13~sec) and different directions of the magnetic field ($\cos\beta = -1, 0, 1$).  
When $\mathcal{F}_r (B, \cos\beta)$ is negative, the momentum transferred from neutrinos and antineutrinos to the matter in the radial direction is suppressed by the magnetic field. 
In general, this is a problem for the SN explosion mechanism which requires an increase of the matter heating and a motion of the matter outwards.

\section{Conclusions}
\label{sec:conclusions}

An influence of a magnetic field on beta-processes is investigated under conditions of the core-collapse supernova. The beta-processes are the dominant channels of the energy exchange between neutrinos and a supernova matter. For study the influence of the magnetic field on the beta-processes, one should consider macroscopic quantities like reaction rates as well as the energy and momentum transferred from neutrinos and antineutrinos to the matter. 
We obtain simple analytical expressions for these quantities for any realistic magnetic field strength known or believed to exist in astrophysical objects. Our results generalize
the ones calculated earlier for an unmagnetized matter and for a magnetized matter transparent for neutrinos and antineutrinos. The matter parameters used in this analysis correspond to the conditions of the supernova region where neutrino can affect the matter and influence a supernova dynamics. The results obtained are applicable for different astrophysical objects, for example, for accretion discs formed at a merger of compact objects in close binary systems.

The magnetic field strength enters the reaction rates and other quantities calculated for the beta-processes through the dimensionless parameters defined in (\ref{eq:par-m-f}). In addition to the strength, they also include the matter parameters~--- the average energy of electron-positron plasma and degree of the lepton degeneracy. These parameters,~$\eta$ and~$\bar\eta$, increase with a growth of the magnetic field strength and the degeneracy of leptons while the increase of an average energy of the electron-positron plasma reduces their values.

Theoretical expressions depend on too many parameters characterizing a matter and neutrino radiation, and this makes a problem to be quite involve for study. Likely, it is possible to fix some of them by using the data from numerical simulations. For this purpose, we have used the results of the 1D PROMETHEUS-VERTEX simulations~\cite{Hudepohl:2014}. Based on the data available, it is possible to reduce the existing number of parameters to two only: the distance from the PNS center and time after a bounce. Because the 1D PROMETHEUS-VERTEX simulations~\cite{Hudepohl:2014} do not take into account the influence of the magnetic field, the other two parameters of the problem are the magnetic field strength and angle between the radius-vector of a reference point and field direction. Our numerical analysis confirms that the influence of the magnetic field on the matter and neutrino parameters is insignificant. So, it is a reasonable assumption to utilize the parameters from the 1D PROMETHEUS-VERTEX simulations for  the magnetic-field effects' study. Namely, we get that in the magnetic field with the strength $B \sim 10^{15}$ G, supernova matter modifications caused by neutrinos are, at most, of a few percents only and, as a consequence, the magnetic-field effects can be safely neglected, considering neutrino interaction and propagation in a supernova matter. Note that both the magnetic field strength and its direction in the beta-processes result a comparable effect.

Numerical analysis of the magnetic-field influence on the chemical composition of a matter and the matter heating shows that this influence is more pronounced just   
after a bounce in a region of the stalled shock wave, and in a more later time, when supernova has already cooled down through the neutrino emission. The magnetic field can both suppress and stimulate the matter heating through the beta-processes in a difference to the unmagnetized matter. The presence of magnetic field results into a decrease of the reaction rates responsible for the matter chemical composition and, as a consequence, a neutron production is suppressed by the magnetic field.

The momentum transferred from neutrinos and antineutrinos to the matter can be decomposed into two components. The first one, $\mathcal{F}_r$, is connected with a partially transparent for neutrino matter, while the second one, $\mathcal{F}_B$, is determined by the asymmetry of the neutrino emission and absorption caused by the magnetic field. As analysis shown, the radial component, $\mathcal{F}_r$, is always suppressed by the magnetic field, and this suppression grows up with the time after a bounce. The $\mathcal{F}_B$ component is smaller than $\mathcal{F}_r$ and reaches the largest values in the first few seconds after a bounce in the vicinity of a protoneutron star surface. Despite~$\mathcal{F}_B$ is relatively small, it can modify the supernova dynamics, namely, it causes an additional non-radial motion of the supernova matter. In particular, it leads to a rotational acceleration of the matter in a region filled with the toroidal magnetic field.

\section*{Acknowledgments}
The work is supported by the Russian Science Foundation (Grant No. 18-72-10070).
We are thankful to H.-T.~Janka and his collaborators for providing us with the data of the supernova explosions and for helpful discussions. A.D.~is thankful to the Max Planck Institute for Physics (Munich, Germany) for hospitality and to Georg Raffelt for support and stimulating discussions.
We thank A.\,Ya.~Parkhomenko, A.~Kartavtsev and V.~Kniazevich for their attention to the paper and valuable comments.


\appendix 

\section{Properties of $I_{k,\varkappa}$ and $J_{k,\varkappa}$}
\label{sec:properties-I-and-J}

We consider properties of the functions $I_{k,\varkappa} (\varepsilon_1, b)$ and $J_{k,\varkappa} (\varepsilon_1, b)$ introduced in Eqs.~(\ref{eq:i11}) and~(\ref{eq:i21}), respectively, which absorb an information about the magnetic field strength. At $b \gg 1$ ($B \gg B_e = 4.41 \times 10^{13}$~G), they are substantially simplified: 
\begin{gather}
\begin{aligned}
\label{eq:i11-mod}
I_{k,\varkappa} (\varepsilon_1, b) \approx &\, \varkappa^{-k-3} \, \varphi_k (\eta) 
= \varkappa^{-k-3} \Big \{ \Gamma (k+3,\eta)
\\
& + (\eta^2 / 4) \, \Big [ \Gamma(k+1) - \Gamma(k+1, \eta) \Big ]  \Big \} ,
\end{aligned}
\\
\begin{aligned}
\label{eq:i21-mod}
J_{k,\varkappa}(\varepsilon_1, b) \approx &\, \varkappa^{-k-3} \, \psi_k (\eta) 
= \varkappa^{-k-3} \, (\eta^2 / 4) \, \Gamma (k+1) .
\end{aligned}
\end{gather}
In this limit, the magnetic field strength is entering through the parameter $\eta = \varkappa \left ( m_e/\varepsilon_1 \right ) \sqrt{2b}$ (the same one, $\bar\eta = \bar\varkappa \left ( m_e/\bar\varepsilon_1 \right ) \sqrt{2b}$, exists for the antineutrino).

As analysis shown, parameters~$k$ and~$\varkappa$ increase with a degeneracy of electrons and neutrinos.
There are two opposite effects of the matter influence on~$\eta$. 
From one side,~$\eta$ is suppressed by the ratio $m_e/\varepsilon_1$, therefore, in the magnetic field with a fixed strength, an ultrarelativistic matter decreases the influence of  the field on the beta-processes~(\ref{eq:p1})--(\ref{eq:p4}). From the other side, when leptons become more degenerate, the parameter~$\varkappa$ grows up. Hence, the parameter~$\eta$ increases with a growth of the degeneracy of leptons. As follows from the definition, $\eta$  takes larger values
with a rise of the magnetic field strength. For~$\bar\eta$, the same arguments are valid.  

\begin{figure}[tb]
\includegraphics[width=0.85\columnwidth]{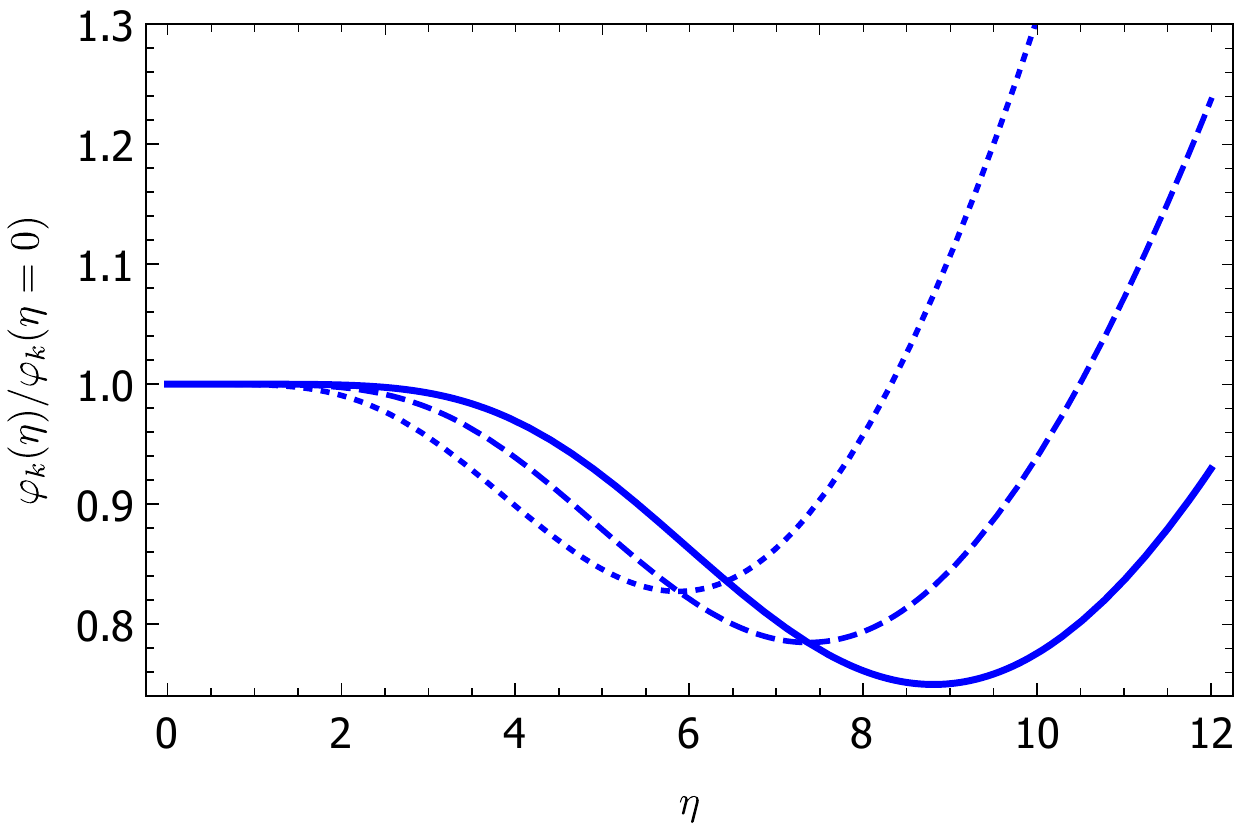}
\caption{The ratio $\varphi_k(\eta)/\varphi_k(\eta = 0)$. 
{\it Dotted line:} $k = 3$. {\it Dashed line:} $k = 4$. {\it Solid line:} $k = 5$.}
\label{fig:plot1}
\end{figure}

In addition, due to the matter electroneutrality, positrons remain non-degenerate everywhere in the supernova. In inner parts of the supernova, electrons become degenerate and, hence, in these regions the influence of the magnetic field on the beta-processes~(\ref{eq:p1}) and (\ref{eq:p2}) with participating of neutrinos is more significant then on the processes~(\ref{eq:p3}) and (\ref{eq:p4}) with antineutrinos. Note that these conclusions coincide with the results of Ref.~\cite{Ognev:2016wlq} obtained for the beta-processes in a matter transparent for neutrinos and antineutrinos.

In Fig.~\ref{fig:plot1}, we plot the function~$\varphi_k (\eta)$~(\ref{eq:i11-mod}) normalized to the reference value~$\varphi_k (\eta = 0)$ in dependence on the parameter~$\eta$ for several typical values~$k$, i.\,e., $k = 3,\, 4,\, 5$. This function has a minimum and, when the value of $k$ increases, the minimum moves to larger values of~$\eta$ and becomes deeper. 

\begin{figure}[tb]
\medskip
\includegraphics[width=0.85\columnwidth]{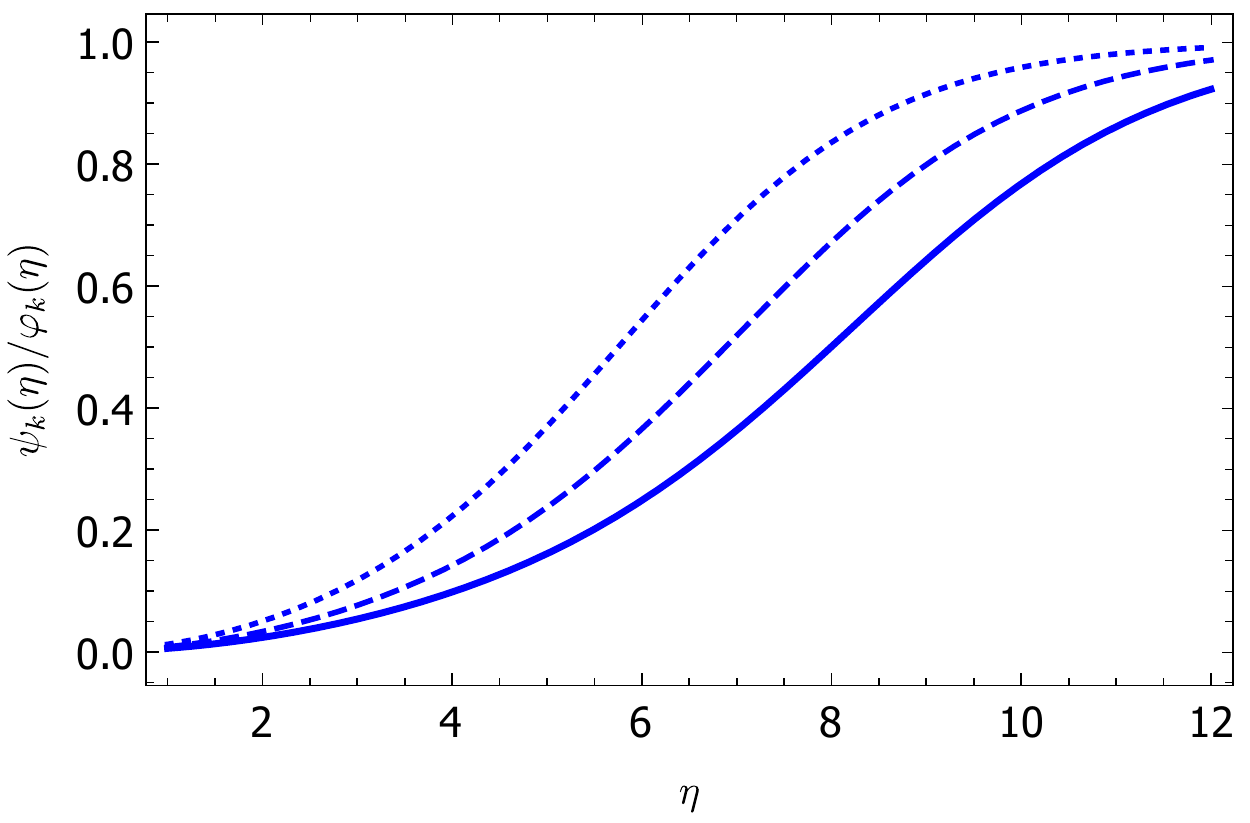}
\caption{The ratio $\psi_k(\eta)/\varphi_k(\eta)$. 
{\it Dotted line:} $k = 3$. {\it Dashed line:} $k = 4$. {\it Solid line:} $k = 5$.}
\label{fig:plot2}
\end{figure}

In Fig.~\ref{fig:plot2}, we present the ratio of the functions $\psi_k (\eta)$~(\ref{eq:i21-mod}) and $\varphi_k(\eta)$~(\ref{eq:i11-mod}) for  $k = 3,\, 4,\, 5$. As seen from Fig.~\ref{fig:plot2}, the function $\varphi_k (\eta)$ reaches $\psi_k (\eta)$ asymptotically at $\eta \gg 1$, because in this limit the incomplete Gamma-functions, $\Gamma(k, \eta)$, in Eq.~(\ref{eq:i11-mod})  turn out to be zero.


\end{document}